\documentclass[twocolumn, aps,pre,superscriptaddress]{revtex4-2}

\usepackage[colorlinks=true,allcolors=blue]{hyperref}
\usepackage{bm}
\usepackage{float}
\usepackage{booktabs,tabularx,array}
\usepackage{graphicx,color}
\usepackage[dvipsnames]{xcolor}

\usepackage{enumerate}
\usepackage{harpoon}
\usepackage{amsmath,amssymb}
\usepackage{gensymb}
\usepackage{accents}
\usepackage[e]{esvect}
\usepackage[type={CC},modifier={by},version={4.0}]{doclicense}
\usepackage{footmisc}

\definecolor{ctcolor}{rgb}{0.8, 0.0, 0.2}

\begin{document}

\preprint{APS/123-QED}

\title{Non-Hermitian chiral surface waves in disordered odd solids}
% \thanks{\doclicenseThis}
\author{Cheng-Tai Lee}
\email[]{chengtailee@tauex.tau.ac.il}
\affiliation{School of Mechanical Engineering, Tel Aviv University, Tel Aviv 69978, Israel}
\author{Tomer Markovich}
\affiliation{School of Mechanical Engineering, Tel Aviv University, Tel Aviv 69978, Israel}
\affiliation{Center for Physics and Chemistry of Living Systems, Tel Aviv University, Tel Aviv 69978, Israel}

% \date{\today}

\begin{abstract}
Chiral surface waves are surface-localized modes that propagate unidirectionally along a boundary, 
enabling directed transport and minimal back-scattering. 
While first identified in quantum systems, they were recently shown to emerge in classical metamaterials in the presence of `odd elasticity'.
Owing to the non-reciprocality
of odd elasticity, these waves exhibit growing amplitudes during propagation, reminiscent of the non-Hermitian skin effect.
To date, studies of odd elastic systems have mainly focused on ordered structures.
Whether structurally-disordered materials %with odd elasticity 
can host non-Hermitian chiral surface waves (NHCSW) remains unexplored.
We address this question using a minimal model of torque-driven disordered odd solids.
Such solids are abundant, from biological gels such as the cytoskeleton driven by motor-proteins to synthesized systems such as magnetic colloidal gels.
We find that torque-driven disordered odd solids have unique NHCSW with stronger surface localization and stable boundary velocity%over a wide range of odd elastic modulus
, in contrast to previous lattice models of odd solids.
%
%We discover a distinctive type of NHCSW with stronger surface localization and stable boundary velocity%over a wide range of odd elastic modulus
%, in contrast to the previous lattice models.
%
These distinct features stem from an intrinsic interplay between boundary torques and odd elasticity in torque-driven odd solids.
Our results offer a new strategy to control NHCSW using active torques.
\end{abstract}

\maketitle

\section*{Introduction}

Chiral surface waves are surface-localized modes propagate unidirectionally along the boundary while decaying into the bulk. They enable directed transport with strongly-suppressed back-scattering and are tolerant to surface roughness or defects. Chiral surface waves have been identified across diverse physical systems, from electron transport in quantum Hall materials~\cite{klitzing1980,halperin1982,buttiker1988,hatsugai1993,hasan2010}, to equatorial waves~\cite{delplace2017}, vortex matter~\cite{bogatskiy2019} and chiral active fluids~\cite{abanov2018,souslov2019,abanov2018,soni2019,caporusso2024a,caprini2025,marconi2026}, and more recently to displacement waves in metamaterials~\cite{scheibner2020a,zhao2020,chen2021,gao2022,fossati2024,veenstra2025}. In many of these examples, the origins of unidirectionality can be traced to broken time-reversal symmetry together with broken mirror symmetry, realized through `odd' viscosity~\cite{avron1998,banerjee2017,abanov2018,souslov2019,soni2019,han2021,markovich2021,hosaka2023,markovich2024,caporusso2024a,hosaka2024,caprini2025,banerjee2025,marconi2026,markovich2025} in fluids and `odd' elasticity in solids~\cite{scheibner2020,scheibner2020a,zhou2020,chen2021,gao2022,shaat2023,fruchart2023,fossati2024,banerjee2025,lee2025,veenstra2024,veenstra2025,caprini2025,nemeth2025b,engstrom2025}. 

% Odd elasticity corresponds to the component of the elastic  tensor $C_{ijkl}$ that is antisymmetric under exchange of index pairs $(i,j)\leftrightarrow (k,l)$~\cite{scheibner2020,fruchart2023}. This odd modulus non-reciprocally couples two independent shear deformations. For example, in two-dimensions (2D), pure shear stress produces simple shear strain, whereas simple shear stress induces negative pure shear strain. In wave dynamics, this non-reciprocity in shear deformations yields an antisymmetric, \textit{non-Hermitian} coupling between longitudinal and transverse modes. 

Odd elasticity (viscosity) corresponds to the component of the elastic (viscous) tensor $C_{ijkl}$ ($\eta_{ijkl}$) that is antisymmetric under exchange of index pairs $(i,j)\leftrightarrow (k,l)$~\cite{scheibner2020,fruchart2023}. This odd modulus couples unconventionally the two independent shear deformations. For example, in two-dimensions (2D), pure shear stress produces simple shear strain (strain rate for viscosity), whereas simple shear stress induces \textit{negative} pure shear strain (strain rate). Accordingly, this coupling yields an antisymmetric coupling between longitudinal and transverse modes in odd viscous~\cite{markovich2024} fluids and, more intriguingly, results in a \textit{non-Hermitian} coupling in odd elastic solids~\cite{fruchart2023}.

The non-Hermitian character of odd solids enables net work extraction~\cite{scheibner2020,scheibner2020a,fruchart2023,fossati2024}. In odd elastic \textit{lattices} assembled from non-reciprocal springs~\cite{chen2021,veenstra2025}, this can manifest as amplitude amplification of chiral surface waves, offering a route to enhance boundary signals and potentially compensate propagation losses. This is similar to the non-Hermitian skin effect (NHSE)~\cite{scheibner2020a}, with the distinction that here the number of surface modes is sub-extensive rather than extensive. Importantly, such odd-lattice realizations typically rely on macroscopic, highly-ordered artificial building blocks and are not straightforward to realize with current material synthesis strategies. 

Recently~\cite{lee2025}, we showed that odd elasticity can broadly emerge in \textit{structurally-disordered} chiral active materials~\cite{furthauer2012,liebchen2022,bowick2022,shankar2022,markovich2019}. We found that the essential requirement is some local injection of active torques to drive internal particle rotations and induce geometric nonlinearities in the elastic response. Crucially, these active sources also generate local boundary torques, which significantly affects surface waves propagation. It therefore remains an open question whether such structurally-disordered odd solids can support chiral surface waves, and how these are modified by the boundary torques.

%%%%%% CHANGING TO BE PNAS STYLE
In this work we study the properties of surface  waves in 2D, disordered odd solid created by active torques~\cite{lee2025}. These waves are the non-Hermitian extension of the well-known Rayleigh waves~\cite{LLelastity}.
We start by a brief review of the model introduced in Ref.~\cite{lee2025} (Sec.~\ref{sec:model}). Then, we formulate the surface-wave problem in a semi-infinite plane (Sec.~\ref{sec:rayleigh surface waves}).
%For simplicity, we focus on an incompressible underdamped solid, neglecting friction and viscosity, and assume a constant odd elastic modulus $K^o$ (created from a uniform, time-independent active torque density $\tau^\circ$). 
%
%
%This work addresses these questions by studying Rayleigh surface waves and is organized as follows. We first briefly review the 2D, disordered, isotropic odd solid created by active torques in Ref.~\cite{lee2025} (Sec.~\ref{sec:model}). We then formulate the Rayleigh surface-wave problem in a semi-infinite plane (Sec.~\ref{sec:rayleigh surface waves}). For simplicity, we focus on the underdamped regime, neglect friction and viscosity, and assume a constant odd elastic modulus $K^o$ (created from a uniform, time-independent active torque density $\tau^\circ$), while taking the incompressible limit. 
%
Despite the presence of boundary torques in this structurally-disordered chiral odd solid, we find non-Hermitian chiral surface waves whose amplitude grows along the propagation direction (Sec.~\ref{sec:unidirectional wave}). Compared with previously-studied odd lattices~\cite{scheibner2020,veenstra2025}, these waves exhibit a much weaker amplitude enhancement (by an order of magnitude), stronger surface localization, and a steady propagation velocity that depends on the sign of odd elasticity yet is essentially insensitive to its magnitude. These qualitative differences point to distinct functionalities and potential material uses of odd solids beyond lattice-based realizations. Finally, we analyze the interplay between  boundary torques and odd elasticity (Sec.~\ref{sec: boundary tau effects}). Although their individual contributions oppose each other, the above features of our chiral odd solid cannot be understood as a simple cancellation. Instead, they stem from the proportionality $K^o=\tau^\circ/4$, intrinsic to odd elasticity generated by active torques~\cite{lee2025}.

% This is reminiscent of non-Hermitian skin effects seen in 1D chain with  non-reciprocal interactions,  ~\cite{scheibner2020a,chen2021,zhang2022e,liu2024e,gohsrich2025}

% $\tau = 4 \alpha K^o$ to see the effects of the sign of between $\tau$ and $K^o$.  
 
% Qualitatively, increasing the torque has the effect to shift the $\tau=0$ curve by the replacement $K^o-(\tau/4)$. For $\alpha \neq 1$, the maximum of the enhancement $q_x^\mathrm{max} \approx 0.5$. Only for $\alpha =1$, this maximum decreased dramatically by an order of magnitude.   

% For small active torques, this enhancement scales linearly with odd elasticity, similar to latticed odd solids of non-reciprocal springs~\cite{veenstra2025}. As torque strength increases, the enhancement reaches a peak and then gradually decrease to eventually vanish. 

\section{Disordered chiral odd solid}\label{sec:model}

Consider a material composed of identical complex particles (namely, not point-like). The material is further assumed to be isotropic and homogeneous at large scales yet locally disordered, which would typically be the case for biological gels such as the cytoskeleton~\cite{Broedersz2014}. 

\begin{figure}[t]
	\centering
	\includegraphics[width=8cm]{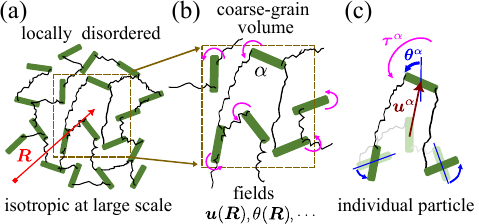}
\caption{(a) Illustration of an elastic material composed of rigid  rod-like particles. Importantly, our model~\cite{lee2025} applies for any other complex rigid particles (granules, colloids, fiber composites, etc.). (b) Coarse-graining at position $\bm{R}$ in the deformed/real space. We consider a locally-disordered, isotropic elastic material, in the presence of local active torque $\tau^\alpha$ ($\alpha$ being the particle index). The various fields ${\bm X}({\bm R})$ are the average of the particle's ${\bm X}^\alpha$ within the coarse-graining volume. (c) Particle displacement $\bm{u}^\alpha$ and internal rotation $\theta^\alpha$ away from the rest position and rest orientation (blue line) in the \textit{undeformed} state. The \textit{individual} rest orientations have no universal direction due to the disordered nature.
}
	\label{fig:CG scheme}
\end{figure}

Local active torques $\tau^\alpha$ (e.g, from actomyosin motor proteins) can be applied at the particle level to induce local \textit{internal} rotation, whose effect is capture in the spirit of the well-known micropolar (Cosserat) elasticity~\cite{eringen1966,eringen1999,eremeyev2013}. The particles are treated as rigid, allowing only translations of center of mass (CM) and rotation around the CM, namely, $\bm{u}^\alpha$ and $\theta^\alpha$ in Fig.~\ref{fig:CG scheme} with $\alpha$ being the particle index. Importantly, both $\bm{u}^\alpha$ and $\theta^\alpha$ cause deformations, which, at the coarse-grained level, gives rise to the usual Cauchy-Green strain and an additional strain due to the internal rotation, see Ref.~\cite{lee2025} for details.

In Ref.~\cite{lee2025} we have shown that by assuming linear stress-strain relations for the two strains~\cite{lee2025}, while retaining geometric nonlinearities in the internal rotation (to leading non-linear order) generated by active torques, odd elasticity naturally emerges.
After elimination of the internal degree-of-freedom $\theta$ that relaxes fast, the Cauchy stress $\bm\sigma$ in {\it real} space position $\bm{R}$ (Eulerian coordinates) takes the form: 
\begin{align}
\sigma_{ij}
=
&
\underbrace{
\frac{\tau}{2}\varepsilon_{ij}
-\kappa \tau^2\delta_{ij}
}_{\bm\sigma^\text{pre}}
% \notag\\
% &+
+\underbrace{
\big[
E_{ijkl}
+\frac{\tau}{4}
(\varepsilon_{jl}\delta_{ik}+\varepsilon_{ik}\delta_{jl})
\big]
}_{\bm{C}}
\nabla_l u_k
\label{eq:Cauchy stress Eule tau}
\, ,
\end{align}
where $\tau$ is the active torque density in the real space $\bm{R}$ and $\nabla_l u_k \equiv \partial u_k/\partial R_l$. The isotropic elasticity tensor is $E_{ijkl}=\lambda\delta_{ij}\delta_{kl} + \mu (\delta_{ik}\delta_{jl}+\delta_{il}\delta_{jk})$ with the Lam\'{e} coefficients $\lambda$ and $\mu$. $\kappa$ is a coefficient that accounts for the energetic cost of rotational mismatch appearing in Cosserat elasticity.

The geometric non-linearity comes to fore in the presence of active torques leading  to the appearance of terms 
${\cal O} \big(\tau\nabla_j u_i\big)$ and ${\cal O} \big(\tau^2\big)$ in the stress.
Notice that the active torques creates a prestress term,
$\bm\sigma^\text{pre}$ in Eq.~\eqref{eq:Cauchy stress Eule tau}, which is non zero even in the absence of deformation.
The elasticity tensor $\bm{C}$ in {\it real} space, presented in the basis of orthogonal strain and stress~\cite{scheibner2020,fruchart2023} is then:
%
%
%Besides the usual linear order ${\cal O} \big(\nabla_j u_i\big)$, the presence of active torque $\tau$ allows terms up to order ${\cal O} \big(\tau\nabla_j u_i\big)$ and ${\cal O} \big(\tau^2\big)$ in stress. $\bm\sigma^\text{pre}$ in Eq.~\eqref{eq:Cauchy stress Eule tau} is the prestress due to the active torques, even in the absence of deformation. $\bm{C}$ is the elasticity tensor in {\it real} space, and takes the matrix form using orthogonal basis for strain and stress~\cite{scheibner2020,fruchart2023}:
%
%
\begin{equation}\label{eq:elasticity tensor}
\begin{matrix}
\includegraphics[width=.35\textwidth]{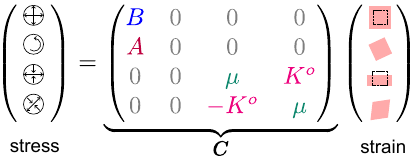}
\end{matrix}
\ \ ,
\end{equation}
where the bulk modulus $B=\lambda+\mu$, $K^o$ is the odd elastic modulus, and the modulus  $A$ couples dilation with torque stress~\footnote{Expressing the elasticity tensor as
$C_{ijkl} = \big[
B\delta_{ij}\delta_{kl}
+
\mu(\delta_{ik}\delta_{jl}+\delta_{il}\delta_{jk}-\delta_{ij}\delta_{kl})
- A\varepsilon_{ij}\delta_{kl}
+K^o(\varepsilon_{ik}\delta_{jl}+\varepsilon_{jl}\delta_{ik})
\big]$ in the 2D irreducible basis gives the matrix ${\bm C}$~\cite{scheibner2020} of Eq.~\eqref{eq:elasticity tensor}. $\delta_{ij}$ and $\varepsilon_{ij}$ are the Kronecker delta and Levi-Civita symbol, respectively.}.
When $\bm\sigma$ is expressed in terms of $\tau$, $A=0$ while $K^o=\tau/4$, as a result of balancing the `total' angular momentum~\cite{markovich2024,lee2025}. This is because $\bm\sigma$ becomes the `total' stress after eliminating the angle variable $\theta$, making $C_{ijkl}$ symmetric under the interchange $i\leftrightarrow j$.  

For practical reasons, it is useful to write $\bm C$ in terms of the torque density in the \textit{undeformed} space $\tau^\circ= \tau(1 + \nabla\cdot\bm{u})$, which is easier to control experimentally~\cite{lee2025}. In this mixed representation (namely, mixing the field in the undeformed space with the real space representation), $A = 2 K^o = \tau^\circ/2$.  We take this case as our model system to examine chiral surface waves. 

\section{Surface Wave Formulation}\label{sec:rayleigh surface waves}

\begin{figure}[t]
	\centering
	\includegraphics[width=6cm]{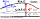}
\caption{Dynamic boundary $h(x,t)$ moves with vertical displacement $u_y(x,0,t)$ (black arrows), measured from the flat undeformed boundary (bottom back dashed). $u_y$ contains a static part $u_y^\text{st}$ (gray dashed arrow) due to static deformations balancing the pre-stress, and a dynamic part $u_y^\text{dyn}$ (red dotted arrow) related to wave excitations. The local normal director $\hat{\bm{N}}$ is determined by the boundary tilt $\nabla_x h = \nabla_x u_y$.%{\color{red}TM: ADD BULK TO THE FIG WITH AXES}
}
	\label{fig:DB}
\end{figure}

We investigate elastic surface waves, which penetrate only to a finite depth from the material boundary. In classical elasticity these are the Rayleigh surface waves~\cite{LLelastity}. 
In general, the form of the surface waves depends on the boundary geometry and can be difficult to solve for arbitrary shapes. Since our goal is to demonstrate the possible existence of chiral surface waves, we consider a simple geometry of a semi-infinite $xy$-plane extending into the bulk along the negative $y$-direction, see Fig.~\ref{fig:DB}.
%
%%% Add IN THE FIG 2
%, i.e., $-\infty < y < 0$ and $|x| < \infty$. 
%In this setup we focus on the regime  close to one boundary $y=0$ and far from the other boundaries, namely, $|x|, |y| \ll \infty$.

Following the standard treatments~\cite{LLelastity, abanov2018, veenstra2025}, we first determine the dispersion relation and the two associated eigenmodes of the bulk (Sec.~\ref{sec:eigenmodes}). We then superpose these two eigenmodes to construct a surface wave that satisfies the stress-free boundary, with \textit{net} zero traction force at the boundary (Sec.~\ref{sec:boundary condition}). This lead to an  equation for $k_x$, whose solutions must describe waves that exist only within the material and decay into the bulk ($y \rightarrow - \infty$) (Sec.~\ref{sec:unidirectional wave}). 

We consider the underdamped case without friction and viscosity, such that elastic waves are sustained by inertia. We begin with a general scenario and subsequently present a practical experimentally realizable example: an odd solid arising from uniform active torques applied in the \textit{undeformed} space, for which $A = 2K^\circ$. For further simplification, we take the incompressible limit, where the bulk modulus $B$ is much larger than other elastic moduli. In this limit, the effect of $A$ modulus is expected to vanish as will be verified below. 

\subsection{Bulk Eigenmodes}\label{sec:eigenmodes}

Elastic waves are governed by the linearized (in terms of $\nabla\bm{u}$) displacement dynamics~\cite{LLelastity,scheibner2020,lee2025}:
\begin{align} \label{eq:displacement dynamic eq}
\rho \ddot{u}_i 
=
\nabla_j \sigma_{ij}
=
\nabla_j(C_{ijkl}\nabla_l u_k )
\, .
\end{align}
Here $\rho$ is the mass density, which is constnat in the homogeneous case we consider. The elasticity tensor $\bm{C}$, given in Eq.~\eqref{eq:elasticity tensor}, includes the odd elastic moduli $K^o$ and $A$ in addition to the bulk and shear moduli $B$ and $\mu$. the prestress $\boldsymbol{\sigma}^{\text{pre}}$ does not appear in Eq.~\eqref{eq:displacement dynamic eq} because in the uniform torque case $\nabla_j \sigma^\text{pre}_{ij} \propto \nabla_j \tau^\circ = 0$. However, the prestress effect becomes prominant when we deal with the surface dynamics below. 

To find the dispersion relation, we use the Fourier transform $\bm{u} = \int \bar{\bm{u}} e^{i(\bm{k}\cdot\bm{R}-\omega t)} {\rm d}{\bm k}/2\pi$
% \ct{[I change the factor from (2\pi)^2 to 2\pi]}
in Eq.~\eqref{eq:displacement dynamic eq}, where $\bar{\bm{u}}$ are the wave amplitudes in the Fourier space, $\bm{k}$ is the wave vector, $\omega$ is the angular frequency, and $\bm{R}\equiv(x,y)$ is the wave position. The amplitudes $\bar{\bm{u}}$ can be decomposed into longitudinal and transverse directions via:
%
%insert the wave ansatz $\bm{u} =\bar{\bm{u}} e^{i(\bm{k}\cdot\bm{R}-\omega t)}$ into Eq.~\eqref{eq:displacement dynamic eq}, where $\bar{\bm{u}}$ are the wave amplitudes, $\bm{k}$ is the wave vector, $\omega$ is the angular frequency and the position $\bm{R}\equiv(x,y)$. The amplitudes $\bar{\bm{u}}$ can be decomposed into the longitudinal and transverse directions via:
%
%
$\bar{u}_{\text{L}} = \hat{k}_i  \bar{u}_i $ and $\bar{u}_{\text{T}} = \varepsilon_{ij}\hat{k}_i\bar{u}_j$, where  
$\bar{u}_\mathrm{L}$ and $\bar{u}_\mathrm{T}$ are their respective amplitudes. $\varepsilon_{ij}$ is the 2D Levi-Civita symbol and $\hat{\bm{k}}\equiv \bm{k}/k $ is the unit wave director with $k \equiv |\bm{k}|$. The dynamic equations are then written as:
\begin{equation}\label{eq:dynamic matrix}
\begin{pmatrix}
B+\mu -({\rho \omega^2}/{k^2}) & K^o
\\    
-K^o+A & \mu-({\rho \omega^2}/{k^2})
\end{pmatrix}
\begin{pmatrix}
\bar{u}_\mathrm{L}
\\
\bar{u}_\mathrm{T}
\end{pmatrix}
=
0
\, ,
\end{equation}
where the off-diagonal terms couple longitudinal and transverse waves. In general, this dynamical matrix is non-symmetric and contains a \textit{non-Hermitian} part, which allows net work extraction from active sources~\cite{scheibner2020a}. 

Solving Eq.~\eqref{eq:dynamic matrix} gives the amplitude eigenmodes:
\begin{align}\label{eq:eigenmodes}
\{
\bar{u}_\text{L}^{(n)} ,\bar{u}_\text{T}^{(n)}
\}
=
\mathcal{N}
\bigg\{
\frac{1}{2}(B + \alpha_n^0 \sqrt{B^2+4Q})
, 
A-K^o
\bigg\}
\, ,
\end{align}
where $n$ refers to the two eigenmodes, one with $\alpha_1^0=1$ and the other with $\alpha_2^0=-1$.
%{\color{red}TM: maybe not $s_n^0$}
Here $\mathcal{N}$ is the normalization factor such that  $(\bar{u}_\text{L}^{(n)})^2 +(\bar{u}_\text{T}^{(n)})^2=1$ and $Q \equiv (A-K^o)K^o$. The dispersion relations gives the corresponding eigenvalues, that we express for later convenience as:
\begin{equation}\label{eq:dispersion r}
r_n 
\equiv
\frac{k_{y}^{(n)}}{k_x}
=
s_n \sqrt{\frac{2\rho \omega^2}{k_x^2\big(
B+2\mu + s^0_n \sqrt{B^2+4Q}
\big)} -1
}
\, .
\end{equation}
The additional sign $s_n= \pm 1$ gives two opposite propagation directions along the $y$-axis. For  surface waves, the proper sign is selected by allowing propagation only within the material, see  Eq.~\eqref{eq:kx eq squared} below.

% \ct{[should comment on bulk-boundary correspondence here?; linear term, e.g., due to anisotropy]}

\subsection{Dynamic Stress-Free Boundary}\label{sec:boundary condition}

Let the odd solid be placed on the $xy$-plane, perpendicular to the direction of gravity, such that gravitational effects may be ignored. The ambient pressure acting on the material boundary is neglected, as it is typically much weaker than the elastic response. Since there is no \textit{external stress}, the material boundary must balance the prestress through elastic deformation, so as to be \textit{stress-free} with zero net traction force acting at it. To formulate this requirement, we need to first describe how the material boundary evolves.

Before any deformation, the material boundary is flat and lies along $y=0$. Due to the presence of time-independent prestress, the odd solid undergoes static deformation through the static displacement $\bm{u}^\text{st}$. As waves are excited, the dynamic displacement $\bm{u}^\text{dyn}$, together with $\bm{u}^\text{st}$, gives the total displacement $\bm{u} = \bm{u}^\text{st} + \bm{u}^\text{dyn}$ (Fig.~\ref{fig:DB}).
The boundary evolution is expressed using the height function $h(x,t)$ that follows particle movement at the boundary: 
$dh/dt = v_y(x, y=h, t)$, with
the material derivative $d()/dt$ and $v_y = d u_y/dt$~\cite{ganeshan2017,abanov2018}. Such framework is often termed `dynamic boundary'~\cite{abanov2018}. 
The boundary is then directly determined by the vertical displacement $h(x,t) = u_y(x, y=h, t) \approx u_y(x, y=0, t)$, which holds to linear order in $|\nabla_j u_i| \ll 1$ (Fig.~\ref{fig:DB}).
%
%Since $v_y = d u_y/dt$~\cite{lee2025}, the boundary is directly determined by the vertical displacement  $h(x,t) = u_y(x, y=h, t) \approx u_y(x, y=0, t)$ within  linear elasticity [Fig.~\ref{fig:DB}], where $|\nabla_j u_i| \sim |\bm{u}| \ll 1$.
%
Accordingly, the stress-free condition at the boundary is written as:
\begin{align}\label{eq:zero traction}
\bm{\sigma}\cdot\hat{\bm{N}} 
=
(\bm\sigma^\text{pre}_{ij} + C_{ijkl}\nabla_l u_k)\cdot \hat{N}_{j}
=
0
\hspace{15pt} \text{at}\hspace{5pt} y=h
\, ,
\end{align}
where $\bm{\sigma}$ is the Cauchy stress tensor of Eq.~\eqref{eq:Cauchy stress Eule tau}, and 
$\hat{\bm{N}} = (-\nabla_x h\ \hat{\bm{x}} + \hat{\bm{y}})/\sqrt{1+(\nabla_x h)^2}$ 
is the unit vector normal to the boundary. 

Ordinarily, in the absence of prestress, $\bm\sigma^\text{pre}$, one ignores nonlinear contributions from the boundary tilt $\nabla_x h$ in $\hat{\bm{N}}$. This gives $C_{ijkl}(\nabla_l u_k)\hat{N}_j \sim |\nabla_j u_i|^2$ to leading order, which effectively treat the boundary as flat, with $h,\nabla_x h \approx 0$ and $\hat{\bm{N}} \approx \hat{\bm{y}}$, reducing  Eq.~\eqref{eq:zero traction} to $\sigma_{xy}=\sigma_{yy}=0$~\cite{abanov2018,veenstra2025}. 
However, this flat-boundary approximation fails in the presence of prestress~\cite{leo1989,huang2025d}. In our odd solid, the pre-torque $(\varepsilon_{ij}\tau^\circ/2)$ couples with the tilt $\nabla_x h = \nabla_x u_y$, creating a non-negligible term of order $\mathcal{O}(\tau^\circ \nabla_j u_i)$, while the contribution from the prepressure $\sim \mathcal{O}\left((\tau^\circ)^2\nabla_j u_i\right)$ can be ignored. 

Substituting Eq.~\eqref{eq:Cauchy stress Eule tau} into Eq.~\eqref{eq:zero traction} yields:
\begin{align} \label{eq:stress-free BC}
&
\begin{pmatrix}
{\tau^\circ}/{2}
\\
{\kappa\big(\tau^\circ\big)^2
% (\lambda+\mu-\kappa_c)}/{4\kappa^2_c
}
\end{pmatrix}
+ 
\bm{M}
\cdot
\bm{d}
% \begin{pmatrix}
% \nabla_x u_x
% \\
% \nabla_x u_y
% \\
% \nabla_y u_x
% \\
% \nabla_y u_y
% \end{pmatrix}
=
0
\, ,
\end{align}
where the leftmost term originates from the static prestress acting on the boundary. We define the displacement-gradient vector $\bm{d} \equiv (\nabla_x u_x,\, \nabla_x u_y,\, \nabla_y u_x,\, \nabla_y u_y)=\bm{d}^\text{st} + \bm{d}^\text{dyn}$, 
with $\bm{d}^\text{st}$ and $\bm{d}^\text{dyn}$
correspond to the gradients of the $\bm{u}^\text{st}$ and $\bm{u}^\text{dyn}$, respectively.
%
%composed of the static  $\bm{d}^\text{st}$ and dynamic $ \bm{d}^\text{dyn}$ parts from $\bm{u}^\text{st}$ and $\bm{u}^\text{dyn}$, respectively. 
The matrix $\bm{M}$ encodes the elastic response and includes a term that arises due to coupling between the pre-torque and the boundary tilt:
\begin{equation}\label{eq:M matrix}
\bm{M}
=
\begin{pmatrix}
-(K^o+A)
& \mu
& \mu
& K^o-A
\\    
B-\mu 
& ({\tau^\circ}/{2}) - K^o
& -K^o
& B+\mu
\end{pmatrix}    
\, .
\end{equation}
Within the linearization scheme, the static part $\bm{d}^\text{st}$ Eq.~\eqref{eq:stress-free BC} neutralize the prestress (see SI~\footnote{See the supplementary materials.}), such that the dynamic part $\bm{d}^\text{dyn}$ that is related to surface waves propagation is solved independently
\begin{align}\label{eq:stress-free BC dynamic part}
\bm{M}\cdot\bm{d}^\text{dyn} = 0 \, .
\end{align}
%
%We leave the static solution, subjected to the full Eq.~\eqref{eq:stress-free BC} with $\bm{d}$ replaced by $\bm{d}^\text{st}$ there, in Appendix~\ref{app:static solution}.

Note that gravity and surface tension also couple to the curved dynamic boundary, as in surface gravity waves~\cite{airy1845,stokes1847,Lamb_book,LLfluidMechanics} or capillary waves~\cite{Lamb_book,LLfluidMechanics,crapper1957}. However, the prestress effect ($\sim \nabla{\bm u}$) is distinct from these two common mechanisms. Gravity contributes a term $\rho g u_y$ ($g$ is the gravitational acceleration) that dominates at long wavelengths, while a surface tension $\gamma$ results in a term $\gamma \nabla_x^2 u_y$ that is important at shorter wavelengths. 
%These two effects are in contrast with the first-order contribution from prestress.
%{\color{red}TM:I THINK WE SHOULD ADD THIS IN THE APPENDIX}

\section{Non-Hermitian Chiral surface waves}\label{sec:unidirectional wave}

The solution to Eq.~\eqref{eq:stress-free BC dynamic part} must be a linear combination of the two eigenmodes found in Eqs.~\eqref{eq:eigenmodes}-\eqref{eq:dispersion r}. Therefore, the surface waves can be written as:
\begin{align}\label{eq:displacemnet ansatz}
\bm{u}^\text{dyn}
% &=u_x \hat{x} + u_y\hat{y}
% \notag
% \\
&= 
\sum_{n=1}^2 m_n \
e^{i\big[k_x(x + r_n y)-\omega t\big]}
\
\big(
\ \bar{u}_\text{T}^{(n)}\hat{\bm{T}}^{(n)}
+ 
\bar{u}_\text{L}^{(n)}\hat{\bm{L}}^{(n)}
\ \big)
\, ,
\end{align}
where the index $n$ in the superscript or subscript refers to the two eigenmodes and $m_n$ are the superposition coefficients. Here $\hat{\bm{L}}^{(n)} \equiv (k_x\hat{\bm{x}}+k_y^{(n)}\hat{\bm{y}})/k$ and $\hat{\bm{T}}^{(n)} \equiv (-k_y^{(n)}\hat{\bm{x}}+k_x\hat{\bm{y}})/k$ are the unit vectors for the longitudinal and transverse directions, respectively, with $k_y^{(n)} \equiv r_n k_x$ (see Eq.~\eqref{eq:dispersion r}). 
Physically, in order to satisfy the stress-free boundary condition, the surface wave of Eq.~\eqref{eq:displacemnet ansatz} combines two modes with different bulk penetration depths and velocities (both set by $\omega/\left(r_n k_x\right)$), but with the same propagation velocity along the surface (set by $\omega/k_x$). 

Inserting Eq.~\eqref{eq:displacemnet ansatz} into Eq.~\eqref{eq:stress-free BC dynamic part} we write an equation in terms of the superposition coefficients $\mathbf{m} \equiv  (m_1, m_2)$, which takes the form $\mathbf{U}\cdot\mathbf{m}=0$ with its full expression in SI~\cite{Note2}.
%{\color{red}WHERE U IS...}. 
Here, $y = h \approx 0$ because the correction is of order $\mathcal{O}\left(\left(\nabla{\bm u}\right)^2\right)$.
%
%Inserting Eq.~\eqref{eq:displacemnet ansatz} into Eq.~\eqref{eq:stress-free BC dynamic part} we get an equation of the from $\mathbf{U}\cdot\mathbf{m}=0$ {\color{red}WHERE U IS...}. Here, $\mathbf{m} \equiv  (m_1, m_2)$ and $y = h \approx 0$ because the leading-order nonlinear effect from $h$ only enters through the coupling between $\tau^\circ$ and the tilt $\nabla_x h$ and is already accounted fot in $\bm{M}$. {\color{red}TM:CHECK}
%
A nontrivial solution requires $\det|\mathbf{U}| = 0$, which gives four equations, each for different $s_n$ choice that gives a possible $\tilde{k}_x$ solution:
\begin{align}\label{eq:kx gen}
4 
+ 
\tilde{k}_x^2\big[
&\tilde{k}_x^2 \phi-2 (2- i s_1\Delta\tilde{\tau})
\big]
\notag\\
&+
s_2 \tilde{k}^2_x \big[
s_1 i \phi \tilde{k}_x^2 -2 \tilde{K}^o
\big]\sqrt{-1+({4}/{\tilde{k}_x^2})}
=
0 
\, ,
\end{align}
where we introduce the dimensionless variables $\tilde{k}_x  \equiv 2 k_x ({{\mu}/{\rho}\omega^2})^{1/2}$ and $\{ \tilde{K}^o, \tilde\tau\} \equiv \{ K^o, \tau\}/\mu$, and defined $\phi \equiv 1- \tilde{K}^o\Delta\tilde\tau^\circ$ with $\Delta\tilde\tau^\circ \equiv  (\tilde\tau^\circ/4) - \tilde{K}^o$. 
In the above, we already took the incompressible limit $B\rightarrow \infty$ such that the effect of the modulus $A$ vanishes. 
Moreover, in this limit, the dispersion relation of Eq.~\eqref{eq:dispersion r} reduces to $r_1 = s_1 i$ and $r_2 = s_2 [-1+(4/\tilde{k}_x^2)]^{1/2}$.
At this stage we keep $\tau^\circ$ and $\tilde{K}^o$ independent (although both are related in our model) to further explore below possible effects of boundary torques that in principle are independent of od elasticity. 

A proper surface wave must obey Eq.~\eqref{eq:kx gen} and two more requirements: (i) It must exist only within the material. Because at this stage the surface is approximated by $h=y\approx 0$, this essentially requires the velocity component in the $\hat{y}$ direction to be non-positive, leading to $\text{Re}(k_y)\leq0$.
%which means $\text{Re}(r_n \tilde{k}_x)\leq0$. such that  $\text{Re}(k_y)<0$, which means negative $y$-velocity. 
(ii) It must decay exponentially into the bulk, namely, $\text{Im}(\tilde{k}_y)<0$. Here $\tilde{k}_y = r_n \tilde{k}_x$ (see Eq.~\eqref{eq:dispersion r}).

\begin{figure}[t]
	\centering
	\includegraphics[width=8.5cm]{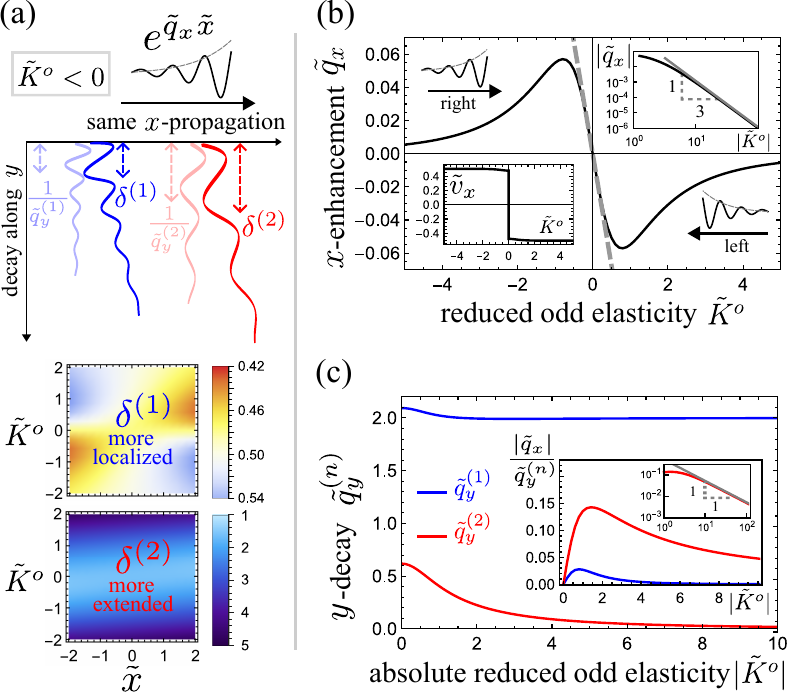}
\caption{(a) (top) Sketch of a non-Hermitian chiral surface wave (for $\tilde{K}^o<0$ as an example) that propagates unidirectionally along the boundary with growing amplitude. The enhancement is $\tilde{q}_x = - {\rm Im}(\tilde{k}_x)$ where $\tilde{k}_x  \equiv 2 k_x ({{\mu}/{\rho}\omega^2})^{1/2}$. This surface wave is comprised of two eigenmodes with the same propagation velocity along the boundary, but with  distinct penetration depths $\delta^{(n)}$. Here, $1/\tilde{q}_y^{(n)}$ are the intrinsic penetration depths. 
(bottom) Color maps of $\delta^{(n)}$ versus $\tilde{K}^o = K^o/\mu$ and the boundary position $\tilde{x} \equiv x({\rho\omega^2/4\mu})^{1/2}$. Along the enhancement direction ($\tilde{q}_x,\tilde{x}>0$ for $\tilde{K}^o<0$ and vice versa), $\delta^{(n)}$ increases and the wave penetrates deeper. %{\color{red}TM:VERIFY FIG}
(b) The enhancement $\tilde{q}_x$ and the wave velocity $\tilde{v}_x = 1/\text{Re}(\tilde{k}_x)$ (bottom inset) versus $\tilde{K}^o$. Notice the  matching sign of $\tilde{q}_x$ and $\tilde{v}_x$ and their behaviors at large $|\tilde{K}^o|$ (insets): vanishing enhancement (after initial linear growth), and almost constant absolute wave velocity. 
(c) Intrinsic inverse  bulk decay lengths $\tilde{q}_y^{(n)}$ of the two modes and their competition with the enhancement, quantified by the localization ratios $|\tilde{q}_x|/\tilde{q}_y^{(n)}$ (inset). The bulk decays always dominates over the enhancement, leading to small and vanishing $|\tilde{q}_x|/\tilde{q}_y^{(n)}$ at large $|\tilde{K}^o|$.   
}
	\label{fig:qx}
\end{figure}

\subsection*{Disordered Odd Solids}

For our disordered chiral odd solids, setting $\tilde{\tau}^\circ = 4\tilde{K}^o$ (equivalently, $\phi=1$ and $\Delta\tilde{\tau}^\circ=0$) and squaring Eq.~\eqref{eq:kx gen} to eliminate $s_2$ results in a cubic equation for $\tilde{k}_x^2$:
\begin{align}\label{eq:kx eq squared}
(1+ i s_1 \tilde{K}^o)\tilde{k}_x^6
-\big[6+&(\tilde{K}^o)^2+4i s_1 \tilde{K}^o\big]\tilde{k}_x^4
\notag\\
&+4 \big[ 2+(\tilde{K}^o)^2 \big]\tilde{k}_x^2
-4
=0
\, .
\end{align}
For $\tilde{K}^o=0$, Eq.~\eqref{eq:kx eq squared} yields \textit{two real} solutions, $\tilde{k}_x=\pm 2.09$,
%{\color{red}TM:EXPLAIN IN FOOTNOTE AND CITE}
recovering the two conventional surface waves of a passive isotropic solid, which propagate in opposite $x$ directions. For $\tilde{K}^o \neq 0$, $\tilde{k}_x$ can become \textit{complex} due to the odd coupling $i \tilde{K}^o$, and the imaginary part of $\tilde{k}_x$ indicates an increasing wave amplitudes along $x$-direction, $\tilde{q}_x\equiv -\text{Im}(\tilde{k}_x)$, see Fig.~\ref{fig:qx}(a). 

In Fig.~\ref{fig:qx}(b) we plot the numerical solution of Eq.~\eqref{eq:kx eq squared}. The enhancement $\tilde{q}_x$ is plotted in the main part as function of $\tilde{K}^o$. The real part of $\tilde{k}_x$ gives the wave velocity and is plotted in the bottom inset, indicating unidirectional wave propagation in the direction opposite to the odd elasticity sign~\footnote{Only one solution of Eq.~\eqref{eq:kx eq squared}, in which  $s_1=s_2=\tilde{K}^o/|\tilde{K}^o|$, satisfies the two condition below Eq.~\eqref{eq:kx gen}}.
To further understand how $\tilde{K}^o$ controls the surface waves propagation and enhancement, we derive the asymptotic expressions for $\tilde{k}_x$. For small $\tilde{K}^o$, we find that $\tilde{k}_x \approx -2.09 s + 0.134 \tilde{K}^oi$, with $s\equiv \tilde{K}^o/|\tilde{K}^o|$, whereas for large $\tilde{K}^o$, we have $\tilde{k}_x \sim   \big[-2 + 0.25(\tilde{K}^{o})^{-2}\big]s+ i(\tilde{K}^{o})^{-3}$~\footnote{To obtain this result we first solve Eq.~\eqref{eq:kx eq squared} for $|\tilde{K}^o| \to\infty$, giving $\tilde{k}_x = -2s$. Then, a linear perturbation about this solution is assumed, $\tilde{k}_x = (-2 s + a) + b i$.} (see top inset of Fig.~\ref{fig:qx}(b)).
These asymptotic expressions imply that the absolute enhancement $|\tilde{q}_x|$ is bounded. For small $\tilde{K}^o$ it grows linearly as $\tilde{q}_x = -0.134 \tilde{K}^o$, it reaches a global maximum at $(|\tilde{K}^o|, |\tilde{q}_x|) \approx (0.8,\ 0.057)$, and then decreases asymptotically as $|\tilde{K}^o|^{-3}$ (see Fig.~\ref{fig:qx}(b)). By contrast, the (dimensionless) boundary velocity remains nearly constant with its sign opposite to $\tilde{K}^o$: $\tilde{v}_x \equiv 1/\text{Re}(\tilde{k}_x) \approx -s/2$.

These surface waves are composed of the two eigenmodes, each decays differently into the bulk with distinct {\it intrinsic} penetration depth $1/\tilde{q}_y^{(n)}$, where $\tilde{q}_y^{(n)} \equiv -\text{Im}(\tilde{k}_y^{(n)})$. 
We find that one mode is more localized with an approximately constant $\tilde{q}_y^{(1)} \approx 2$. The other mode extends deeper into the bulk, with $|\tilde{q}_y^{(2)}|$ that decreases from $\sim 0.5$ as $\tilde{K}^o$ increases, see Fig.~\ref{fig:qx}(c). 
Crucially, because of the enhancement $\tilde{q}_x$, the penetration depth become position-dependent and increase as the wave propagate (Fig.~\ref{fig:qx}(a)). 
To quantify the competition between the enhancement and  decay, we define the penetration depths $\delta^{(n)} = 1/\tilde{q}_y^{(n)} + \tilde{x}\tilde{q}_x/\tilde{q}_y^{(n)}$~\footnote{The penetration depth $\delta^{(n)}$ is defined through the decay condition: $\exp\big(\tilde{q}_x \tilde{x}-\tilde{q}_y^{(n)} \delta^{(n)}\big)=\exp(-1)$}, where $\tilde{x} \equiv x({\rho\omega^2/4\mu})^{1/2}$.
The first term is the intrinsic penetration depth, free from the enhancement effect, while in the second term, the absolute ratio $|\tilde{q}_x|/\tilde{q}_y^{(n)}$ captures the competing enhancement and decay (see inset of Fig.~\ref{fig:qx}(c)).

%Owing to the enhancement $\tilde{q}_x$, this surface wave penetrates more deeply and becomes progressively less localized as it propagates [Fig.~\ref{fig:qx}(a)]. To quantify the competition between the enhancement and decay, we define the penetration depths $\delta^{(n)} = 1/\tilde{q}_y^{(n)} + \tilde{x}\tilde{q}_x/\tilde{q}_y^{(n)}$~\footnote{The penetration depth $\delta^{(n)}$ is defined through the decay condition: $\exp\big(\tilde{q}_x \tilde{x}-\tilde{q}_y^{(n)} \delta^{(n)}\big)=\exp(-1)$}, where $\tilde{x} \equiv x({\rho\omega^2/4\mu})^{1/2}$. The first term $1/\tilde{q}_y^{(n)}$ represents the intrinsic penetration depth, free from the enhancement effect. In the second term, the absolute ratio $|\tilde{q}_x|/\tilde{q}_y^{(n)}$ 
%(for along the $x$-propagation direction) 
%captures such effect and determines the material aspect ratio required for the wave to decay before reaching the bottom BD. This ratio thus provides a quantitative criterion for validity of neglecting the bottom BD effects, and also serves as a measure of surface localization across the system -- the smaller, the more localized.

In the conventional treatment of surface waves (and also here) the bottom boundary is ignored as the waves decay long before reaching it. The ratio $|\tilde{q}_x|/\tilde{q}_y^{(n)}$ determines the bulk aspect ratio for which this assumption is valid.
To illustrate this, consider a solid of width $L_x$ and depth $L_y$, with $L_x,L_y \gg 1/\tilde{q}_y^{(n)}$ such that system dimensions are large compare to the intrinsic penetration depth. To ensure the bottom boundary can be ignored, we must have $|\tilde{q}_x| L_x \ll \tilde{q}_y^{(n)} L_y$. In our odd solid, this ratio remains below $\sim 0.15$ for all $\tilde{K}^o$ and decreases asymptotically as $(\tilde{K}^o)^{-1}$ for large $\tilde{K}^o$ (inset of Fig.~\ref{fig:qx}(c)). 
Details on the asymptotic expression can be found in the SI~\cite{Note2}.
%{\color{red}TM: NOT SURE IF TO HAVE SI OR MATERIALS}

Overall, we find the emergence of \textit{non-Hermitian chiral} surface waves, which propagate unidirectionally (opposite to the sign of odd elasticity $\tilde{K}^o$) with enhanced amplitude along their propagation direction. We thus expect these waves to accumulate at the boundary. 
This unidirectional propagation, enhancement, and corner accumulation are reminiscent of the non-Hermitian skin effect (NHSE)~\cite{zhang2022e,liu2024e,gohsrich2025}, which can arise, for instance, in chains or lattices with non-reciprocal site interactions~\cite{ghatak2020,chen2021}, or in anisotropic odd elastic materials~\cite{scheibner2020a}.
%{\color{red}TM:PUT CITATION PER EXAMPLE AND MAYBE A REVIEW IN THE BEGINNING}
However, the non-Hermitian chiral surface waves we find form only a subset of the wave spectrum. Their number scales with the surface boundary size. By contrast, the number of surface states in the NHSE is extensive and scales with the system size. This difference lies in the isotropic bulk dynamic matrix, Eq.~\eqref{eq:dynamic matrix}, that depend on $k^2$ and is thus symmetric under $k_x \leftrightarrow -k_x$ at fixed $k_y$ (and vice versa). NHSE in odd elastic systems requires some form of anisotropy~\cite{scheibner2020a} that breaks this symmetry.

%Consequently, the non-Hermitian bulk-boundary correspondence underlying the NHSE~\cite{scheibner2020a,ghatak2020}, which typically requires terms $\sim k_x$, $k_y$, or $k_xk_y$ in the dynamical matrix, does not occur here.

%%%%%%%

%Overall, our disordered odd solids support unidirectional surface waves with weakly-enhanced propagation at an approximately steady velocity.
%
In our model of disordered odd solid, the effects of boundary torques are coupled to odd elasticity through $\tilde\tau^\circ = 4 \tilde{K}$, making it difficult to disentangle their independent roles in the appearance of non-Hermitian chiral surface waves. In next section, we relax this constraint to examine the distinct effects, and highlight the remarkable features stemming from the specific relation $\tilde\tau^o/\tilde{K}^o = 4$.

\begin{figure}[t]
	\centering
	\includegraphics[width=8.5cm]{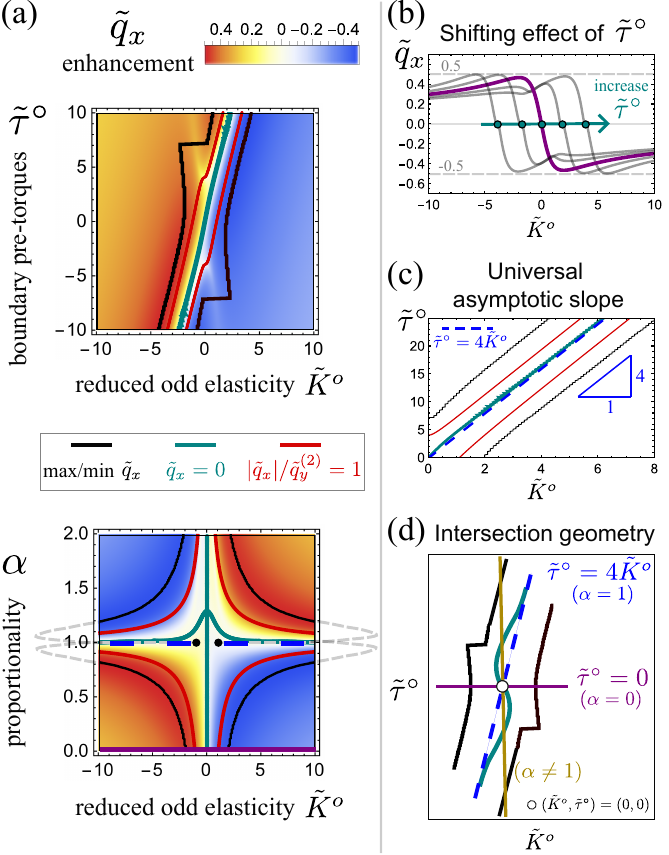}
\caption{(a) Color maps of the enhancement $\tilde{q}_x$ versus odd elasticity $\tilde{K}^o$ and the boundary torque $\tilde\tau^\circ$ (top)/the proportionality $\alpha\equiv \tau^\circ/4\tilde{K}^o$ (bottom). All contour curves use the same color scheme across panels. Note that the contour curves are constructed by scanning over $\tilde\tau^\circ$ (or $\alpha$) [details in the main text].
We first focus on the role of $\tilde\tau^\circ$.
(b) For large $|\tilde\tau^\circ|$, the dominant effect of $\tilde\tau^\circ$ is approximately an horizontal shift of the zero torque curve ($\tilde\tau^\circ=0$, purple; gay from left to right: $\tau^\circ=$-16, -8, 8, 16).  
%{\color{red}TM: ADD $\tau$ VALUES}
% 
(c) The shifting effects of $\tilde\tau^\circ$ leads to the universal asymptotic slope $d\tilde{\tau}^\circ/d\tilde{K}^o=4$ for all the contours at large $\tilde\tau^\circ$. In particular, the contour $\tilde{q}_x=0$ (green) has the asymptote $\tilde\tau^\circ = 4 \tilde{K}^o$, corresponding to our disordered odd solid. (d) A schematic intersection analysis from geometry illustrates the origin of the discrete transition in the max/min contours at $\alpha=1$ (our odd solid, blue dashed) at the bottom of panel a.} 
	\label{fig:tau effects}
\end{figure} 

\section{Boundary Torques Effects}\label{sec: boundary tau effects}

To explore the effects of boundary torques, we treat $\tilde{\tau}^\circ$ and $\tilde{K}^o$ as  independent. Such a scenario may be realized, for example, by incorporating active torques~\cite{lee2025} into odd lattices of non-reciprocal springs~\cite{scheibner2020,fruchart2023}.  

For fixed $\tilde{\tau}^\circ$, varying $\tilde{K}^o$ produces a curve for the enhancement $\tilde{q}_x$, which is similar to that in Fig.~\ref{fig:qx}(b). For each pair $(\tilde{\tau}^\circ,\tilde{K}^o)$ we extract the curve's global maximum, minimum, and the zero of $\tilde{q}_x$, as well as the $\tilde{K}^o$ value for which $|\tilde{q}_x|/\tilde{q}_y^{(n)} = 1$ as a measure of the extent of surface localization. Tracing these characteristic points yields the contours overlaid on the $\tilde{q}_x$ color map in the $(\tilde{\tau}^\circ,\tilde{K}^o)$ space, see Fig.~\ref{fig:tau effects}(a, top). 
The resulting map has a few generic features. First, across all explored $\tilde{\tau}^\circ$, the absolute maximum and minimum of $\tilde{q}_x$ are practically constant with value $\approx 0.5$ (along the black curves). Second, increasing $|\tilde{K}^o|$ at fixed $\tilde\tau^\circ$ we find that $|\tilde{q}_x|/\tilde{q}_y^{(1)}$ remains always below unity, whereas $|\tilde{q}_x|/\tilde{q}_y^{(2)}$ can become larger for values below (above) the bottom (top) red curve, indicating weaker surface localization at larger $|\tilde{K}^o|$. %{\color{red}TM:NOT CLEAR}
Remarkably, an exception for the above generic features is on the line $\tilde{\tau}^\circ=4\tilde{K}^o$, corresponding to our model odd solid. Note that at the scale of the color map it is essentially on top of the turquoise line, see Fig.~\ref{fig:tau effects}(d) for a qualitative illustration). Along this line the maximal enhancement is an order of magnitude weaker ($\sim0.057$), the surface localization is stronger with $|\tilde{q}_x|/\tilde{q}_y^{(n)}<1$ for all $\tilde{K}^o$, and the boundary velocity is almost constant $|v_x|\approx 1/2$ (Fig.~\ref{fig:qx}(b)).
%{\color{red}TM:WHAT IS THE VELOCITY FOR OTHER VALUES? WE NEVER SHOW/DISCUSS THIS}

Importantly, the boundary torques $\tilde\tau^\circ$ acts oppositely to odd elasticity $\tilde{K}^o$. Qualitatively, this can be understood as follows. In the absence of odd elasticity, for $\tilde\tau^\circ>0$  we find that $\tilde{q}_x>0$, while for $\tilde\tau^\circ=0$, to get $\tilde{q}_x>0$ we must have $\tilde{K}^o<0$. When both $\tilde\tau^\circ$ and $\tilde{K}^o$ are present, their contributions cancel off on the contour of zero $\tilde{q}_x$ (turquoise contour), in which case the odd solid is effectively passive, allowing propagation of classical Rayleigh surface waves in both directions~\footnote{This is because the stress-free boundary equations are functions of $\tilde{k}_x^2$ such that their real non-zero solutions must come in pairs.}. Hence, the contour $\tilde{q}_x=0$ also marks the reversal of the boundary propagation direction.
The detailed effects of $\tilde\tau^\circ$ has two regimes, separated by the abrupt horizontal jumps in the max/min $\tilde{q}_x$ contours (black line in Fig.~\ref{fig:tau effects}(a)). For $|\tilde\tau^\circ| \lesssim 7.5$, the dependence of $\tilde{q}_x$ on $\tilde{K}^o$ is intricate and is related to a competition between two local extrema (see the figure in the SI~\cite{Note2}). At $|\tilde\tau^\circ| \approx 7.5$ a different local exterema becomes the global one, resulting in a plateau in the black contour, see details in the SI~\cite{Note2}.
%non-trivial, {\color{red}because the two local peaks (or valleys) compete to become the global extremum, see Appendix~\ref{app:support plots}TM:NOT CLEAR}. 
For $|\tilde\tau^\circ| \gtrsim 7.5$, the dominant effect is a shift of the $\tilde\tau^\circ=0$ curve 
%by an amount $\approx \tilde\tau^\circ/4$ 
along the $\tilde{K}^o$-axis, see Fig.~\ref{fig:tau effects}(b). This implies a universal asymptotic behavior at large $\tilde\tau^\circ$ for all  contours. We numerically find such universal scaling with slope $d \tilde\tau^\circ/d\tilde{K}^o =4$, see Fig.~\ref{fig:tau effects}(c). 

The proportionality $\tilde{\tau}^\circ=4\tilde{K}^o$ we have in our model odd solid, together with the universal asymptotic slope $d\tilde{\tau}^\circ/d\tilde{K}^o=4$ we find numerically, underlies the  distinctive features of our model (weak enhancement, strong localization, and constant velocity).
%
%
%Remarkably, the proportionality $\tilde{\tau}^\circ=4\tilde{K}^o$ we have in our model odd solid, together with the universal asymptotic slope $d\tilde{\tau}^\circ/d\tilde{K}^o=4$, underlies the  distinctive features of our model: nearly an order of magnitude weaker maximal enhancement ($\sim0.057$), stronger surface localization with $|\tilde{q}_x|/\tilde{q}_y^{(n)}<1$ for all $\tilde{K}^o$, and an almost constant boundary velocity $|v_x|\approx 1/2$ (Fig.~\ref{fig:qx}(b)).
%{\color{red}TM:WHAT IS THE VELOCITY FOR OTHER VALUES? WE NEVER SHOW/DISCUSS THIS}
%
The line $\tilde{\tau}^\circ=4\tilde{K}^o$ is unique in the $(\tilde{\tau}^\circ,\tilde{K}^o)$ space because it is parallel to the large-$\tilde{\tau}^\circ$ asymptotes of the contours (Fig.~\ref{fig:tau effects}(c)) and does not intersect the max/min $\tilde{q}_x$ curves or the contour $|\tilde{q}_x|/\tilde{q}_y^{(n)}=1$. Instead, it stays close to the zero $\tilde{q}_x$ contour where the chiral effects from odd elasticity and boundary torques cancel each other,  Fig.~\ref{fig:tau effects}(d). Accordingly, the  surface dynamics for that case differs only slightly from those of conventional non-chiral solids, while still retaining its chiral nature. 
%This results in a persistently small $|\tilde{q}_x|$ and an approximately constant $|\tilde{v}_x|$ for all $\tilde{K}^o$.

To illustrate the distinctive features of our model disordered odd solid, we plot in Fig.~\ref{fig:tau effects}(a, bottom) a color map of the ratio $\alpha\equiv \tilde{\tau}/(4\tilde{K}^o$ and $\tilde{K}^o)$, where $\alpha$ is the ratio of boundary torques and odd elasticity normalized to our model for which $\alpha=1$. 
As $\alpha$ starts deviating from unity, the corresponding line $\tilde{\tau}=4\alpha \tilde{K}^o$ intersects the max/min $\tilde{q}_x$ (black) contours at  infinitely large $|\tilde{K}^o|$ and approaches $\tilde{K}^o=0$ with increased deviation. This is depicted qualitatively in Fig.~\ref{fig:tau effects}(d). 
The existence of such intersection provides a qualitative explanation for the discrete jumps of the max/min $\tilde{q}_x$ contour as $\alpha\to 1$ in Fig.~\ref{fig:tau effects}(a, bottom).
Note that the two dots at $\alpha=1$ are fundamentally different from these intersections, and exhibit a much smaller $\tilde{q}_x$ as discussed earlier. Similar analysis also explains the appearance of three zeros for $\tilde{q}_x$ in the regime $1<\alpha<1.25$.

%Next, we illustrate the distinctive `transition' behavior of our disordered odd solid. To this end, we vary the ratio between the BD torque and odd elasticity, $\alpha\equiv \tilde{\tau}/(4\tilde{K}^o)$. As $\alpha$ starts to deviate from our odd solid case $\alpha=1$, the corresponding line $\tilde{\tau}=4\alpha \tilde{K}^o$ intersects the max/min $\tilde{q}_x$ contours from infinitely large $|\tilde{K}^o|$ to gradually approach $\tilde{K}^o=0$ with increased deviation [Fig.~\ref{fig:tau effects}(d)]. This intersection behavior qualitatively provides a geometric explanation for the discrete jumps of the max/min $\tilde{q}_x$ contour as $\alpha\to 1$ in the $\tilde{q}_x$ color map obtained by scanning $\alpha$ [Fig.~\ref{fig:tau effects}(a,bottom)].
% Here the contours are constructed using the same procedure as in Fig.~\ref{fig:tau effects}(a,top). 
%Note that the two dots at $\alpha=1$ are fundamentally different from these intersections, and exhibit a much smaller $\tilde{q}_x$ as discussed earlier. At last, the similar interaction analysis also explains multiple zero-crossings in the regime $1<\alpha<1.25$.

% These analyses together show that our disordered odd solid from active torques ($\alpha=1$, equivalently, $\tilde\tau^\circ= 4 \tilde{K}^o$) represents a distinctive limit relative to other odd-lattice realizations with or without adding constant BD torques.

\section{Discussion and Conclusion}

In conventional isotropic passive solids, (Rayleigh) surface waves propagate in both directions. With the introduction of odd elasticity, surface waves become chiral and non-Hermitian, with amplitudes growing while propagating unidirectionally. There are two mechanisms to produce such waves: the non-Hermitian skin effect (NHSE)~\cite{ghatak2020,scheibner2020a,chen2021,zhang2022e,liu2024e,gohsrich2025}, and non-Hermitian chiral (Rayleigh-like) surface waves (NHCSW). The key difference between these mechanisms lies in the structure of the bulk dynamical matrix, which in turn leads to different mode counting. The number of such surface modes scales with boundary size for NHCSW and with the entire system size for NHSE. In odd elastic lattices using engineered nonreciprocal springs~\cite{scheibner2020,fruchart2023}, both types were reported depending on their detailed designs (e.g., NHSE~\cite{scheibner2020a,chen2021} and NHCSW~\cite{veenstra2025}).  

This work shows that structurally-disordered odd solids, in which odd elasticity stems from active torques (Fig.~\ref{fig:CG scheme}), can host NHCSW (Fig.~\ref{fig:qx}). 
We have studied the incompressible, underdamped limit. Nevertheless, finite bulk modulus and viscosity may modify the dispersion relation, allowing the system to pass beyond the `exceptional point', 
%(making $B^2+4Q\leq 0$ in Eqs.~\eqref{eq:eigenmodes} and \eqref{eq:dispersion r}) and 
thereby inducing new dynamics~\cite{gao2022,lee2025}. This will be explored in future work.
Such torque-driven odd solids are expected to broadly arise in biological or synthesized systems with examples including cytosketal networks driven by motor proteins and magnetic colloidal gels. Unlike odd elastic lattices, these system feature non-negligible boundary-torque effects that can greatly modify the behavior of NHCSW.

Boundary torques typically induce chiral effects opposite to those of odd elasticity. Notably, this counteraction, together with the proportionality $\tau^\circ=4K^o$ intrinsic to our torque-driven odd solids, results in their unique and distinctive features. More specifically, their surface dynamics differs only slightly from that of conventional passive solids while still retaining their chiral nature. This leads to stronger surface localization, weaker amplitude enhancement and an almost constant boundary velocity over a wide range of $K^o$. 
In stark contrast, NHCSW in odd elastic lattices ($\tau^\circ =0$)~\cite{scheibner2020,veenstra2025} and odd solids with different $(\tau^\circ$,$K^o)$ such that $\tau^\circ/4K^o \neq 1$, are sensitive to increasing $|K^o|$, with faster boundary velocity and stronger enhancement, which leads to weaker surface localization (Fig.~\ref{fig:tau effects}) and ultimately more pronounced corner accumulation.

These boundary-torques features mark distinct design purposes of odd elastic materials. Odd elastic lattices~\cite{scheibner2020} may be suited for faster transport and strong signal enhancement. On the other hand, the disordered odd solids studied here are suitable for surface-localized transport where amplitude growth is undesirable, as it can trigger nonlinear, potentially destabilizing dynamics~\cite{jana2025b}. 
Our work allows for other intermediate designs using a combination of active torques within odd lattices, to control NHCSW features.
Given the relevance of disordered odd solids to biological gels~\cite{markovich2019,furthauer2012}, and the possibilities of controlling edge-transport properties in metamaterials, it would be interesting to further explore NHCSW and their potential roles.

\begin{acknowledgments}

This research was supported in part by Grant No. 2022/369 from the United States-Israel Binational Science Foundation (BSF). T.M. acknowledges funding from the Israel Science Foundation (Grant No. 1356/22). 

% Following the \href{https://www.coalition-s.org/rights-retention-strategy/}{Rights Retention Strategy of Plan S}, a CC-BY 4.0 public copyright license has been applied by the authors to the present document and will be applied to all subsequent versions up to the Author Accepted Manuscript arising from this submission.
\end{acknowledgments}

\bibliographystyle{rsc}
\bibliographystyle{apsrev4-1}
\bibliography{references}

\end{document}